\documentclass{ws-p8-50x6-00}

\newcommand{\msbar}{\mbox{\tiny $\overline{MS}$}}        

\def\lsim{\mathrel{\rlap{\lower4pt\hbox{\hskip1pt$\sim$}}
    \raise1pt\hbox{$<$}}}                
\def\gsim{\mathrel{\rlap{\lower4pt\hbox{\hskip1pt$\sim$}}
    \raise1pt\hbox{$>$}}}                

\def\3{\ss}

\renewcommand{\jobname}{Manuscript}
\renewcommand{\today}{\date{August 10, 2001}}
\renewcommand{\rightnote}{{\it \hspace*{1.3cm} Talk to be published by {\bf World Scientific}
Publishing Co. }}
\renewcommand{\newpage}{}

\begin{document}
\vspace*{-2.5cm}
\begin{flushleft}
{\normalsize DESY 01-110} \hfill\\
{\normalsize August 2001} 
\end{flushleft}
\vspace{0.7cm}

\title{Nucleon form factors and structure functions from Lattice QCD\thanks{
T{\lowercase{alk given by}} G. S{\lowercase{chierholz at}} 
`W{\lowercase{orkshop on}} L{\lowercase{epton}} S{\lowercase{cattering,}} 
H{\lowercase{adrons and}} QCD', 
A{\lowercase{delaide}} U{\lowercase{niversity,}} M{\lowercase{arch}} 26 - 
A{\lowercase{pril}} 6, 2001.}}

\author{M.~G\"ockeler$^1$, \, R.~Horsley$^2$, \, D.~Pleiter$^2$,}
\author{P.~E.~L. Rakow$^1$, \, G. Schierholz$^{2,3}$\\ $\:$}

\address{$^1$Institut f\"ur Theoretische Physik,
         Universit\"at Regensburg, \\ D-93040 Regensburg, Germany}
\address{$^2$John von Neumann-Institut f\"ur Computing NIC,\\ Deutsches 
Elektronen-Synchrotron DESY, D-15735 Zeuthen, Germany}
\address{$^3$Deutsches Elektronen-Synchrotron DESY, D-22603 Hamburg, Germany}


\maketitle

\abstracts{
In this talk we highlight recent lattice calculations of the nucleon
form factors and structure functions.}


\section{Introduction}

The lattice formulation of {\it QCD} is, at present, the only known way of
obtaining low energy properties of the theory in a direct way,
without any model assumptions. Quantities within the grasp of
lattice {\it QCD} involving light quarks include the hadron mass spectrum,
quark masses, the $\Lambda$ parameter, the chiral condensate,
the sigma term, decay constants,
the axial and tensor charge of the nucleon,
form factors and twist two and four polarised and unpolarised
moments of structure functions for the nucleon, pion and rho,
to name only a few of the possibilities. Our group (the {\it QCDSF}
collaboration) has been actively involved for the last few years
in attempting to determine some of these quantities, all characterised
by their non-perturbative nature. Rather than giving an exhaustive progress 
report of recent developments in the field\cite{bangalore00a}%
we shall, due to lack of space, focus our attention on 
two topics: nucleon form factors (including the axial charge)
and moments of structure functions and how the lattice method
can lead to their determination.

The lattice approach involves first euclideanising the {\it QCD}
action and then discretising space-time (with lattice
spacing $a$). The path integral then becomes a very high dimensional
partition function, which is amenable to Monte Carlo methods
of statistical physics. This allows  
correlation functions (which can be related to {\it QCD} matrix elements)
to be determined. Progress in the field is slow.
First our `box' must be large enough to fit our correlation functions into.
Then, often, a chiral extrapolation must be made from a quark mass region
around the strange quark mass to the light up and down quarks.
Finally the continuum limit must be taken, i.e. $a \to 0$.
This is all very time consuming as, in the statistical
mechanics picture, we are approaching a second
order phase transition, with all its attendant problems.
Additionally, simply to save computer time, the fermion determinant
in the action is often discarded. This `quenched'
or `valence' procedure is an uncontrolled approximation.
Recently, however, simulations with two mass-degenerate
sea quarks have begun appearing, allowing a first look at the possible
effects of quenching. Finally, in addition to the above problems,
to be able to compare with phenomenological or experimental results,
matrix elements must be renormalised.
In total this is an ambitious programme.


\section{Generalities}


\subsection{The processes}

Lepton--nucleon elastic scattering, $lN \to lN$, in which a photon
is exchanged between the lepton (usually an electron) and the nucleon
(usually a proton), has been studied for many years.
Indeed there has been a resurgence of interest in these processes
as part of the Jefferson Laboratory (Jlab) `hadron' physics programme.
The scattering matrix element can be decomposed
into a known electromagnetic piece and an unknown {\it QCD}
matrix element with decomposition:
\begin{equation}
  \langle N(\vec{p\:}^\prime,\vec{s\:}^\prime) | J^{\mu}(\vec{q})
                          | N(\vec{p}, \vec{s}) \rangle =
           \overline{u}(\vec{p\:}^\prime,\vec{s\:}^\prime)
              \left[ \gamma^\mu 
                     F_1(Q^2) +
                     i\sigma^{\mu\nu} {q_\nu \over 2m_N}
                     F_2(Q^2)
              \right] u(\vec{p},\vec{s}) \,,
\end{equation}
where $q = p^\prime - p$ is the momentum transfer and $Q^2 = - q^2 > 0$.
The values at $Q^2 =0$ are $F^p_1(0)= 1$ due to the conservation
of the vector current and $F^p_2(0) = \mu^p - 1$, the anomalous
magnetic moment, in units of $e/2m_N$ or magnetons.
($F^n_1(0) = 0$, $F^n_2(0) = \mu^n$.)
Experimentally, it is more convenient to define the Sachs form factors
\begin{eqnarray}
   G_e(Q^2) &=&
      F_1(Q^2) - {Q^2\over (2m_N)^2} F_2(Q^2) \,,
                                                \nonumber \\
   G_m(Q^2) &=& F_1(Q^2) + F_2(Q^2) \,.
                                                \nonumber
\end{eqnarray}
Earlier experimental results give phenomenological (dipole) fits of
\begin{equation}
   G_e^p(Q^2) = {G_m^p(Q^2) \over \mu^p}
              = {G_m^n(Q^2) \over \mu^n}
              = \left(1+ {Q^2\over m_V^2} \right)^{-2} \,, \qquad
   G_e^n(Q^2) = 0 \,,
\label{Gem_phen_fits}
\end{equation}
with $m_V \approx 0.82\, \mbox{GeV}$, $\mu^p \approx 2.79$ and
$\mu^n \approx -1.91$.

Similarly neutrino--nucleon scattering, for example 
$\nu_\mu n \to \mu^- p$ mediated by a $W^+$ exchange,
leads to an unknown axial current
hadronic matrix element between neutron and proton states,
which, with the use of current algebra and isospin invariance,
may be re-written between $p$ states alone and has Lorentz decomposition
\begin{equation}
  \langle p(\vec{p\:}^\prime,\vec{s\:}^\prime) | A^{u-d}_\mu(\vec{q})        
                          | p(\vec{p},\vec{s}) \rangle =
         \overline{u}(\vec{p\:}^\prime,\vec{s\:}^\prime)
              \left[ \gamma_\mu\gamma_5
                    g_A(Q^2) +
                     i\gamma_5 {q_\mu \over 2m_N} h_A(Q^2)
              \right] u(\vec{p},\vec{s})
\end{equation}
where $A^{u-d}_\mu = \overline{u}\gamma_\mu\gamma_5 u -
                     \overline{d}\gamma_\mu\gamma_5 d$.
Experimental results give phenomenological fits
\begin{equation}
   g_A(Q^2) = g_A(0)\left(1+ {Q^2\over m_A^2}
                        \right)^{-2} \,,
\end{equation}
with $m_A \approx 1.00 \,\mbox{GeV}$. From the $\beta$ decay we 
know\cite{groom00a} $g_A \equiv g_A(0) = 1.2670(35)$.

At higher momentum transfer, the nucleon is broken up
by the photon (or $W^{\pm}$) probe. We enter the regime
of {\it deep inelastic scattering} experiments $eN \to e X$
(or $\nu_\mu n \to \mu^- X$). The operator
product expansion, OPE, leads to relations between moments
of the structure functions and certain nucleon matrix elements.
For example,
\begin{equation}
   \int_0^1 dx x^{n-2} F_2(x,Q^2) 
           = {1 \over 3} E^{\msbar}_{n}({\mu^2\over Q^2},g^{\msbar})
                    v_n^{\msbar}(g^{\msbar})
                + O({1\over Q^2}) \,.
\label{OPE_result}
\end{equation}
Here $x$ is the Bjorken variable and
$v_n^{\msbar} \propto \langle N| {\cal O}_n| N \rangle^{\msbar}$,
where ${\cal O}_n \sim \overline{q}\gamma_{\mu_1} D_{\mu_2} \ldots D_{\mu_n}q$
are operators bilinear in the quarks, each containing
$n-1$ covariant derivatives. For $eN$ scattering only even moments
contribute, while for $\nu N$ all moments are allowed.

All the matrix elements briefly discussed above can be determined
non-perturbatively using lattice {\it QCD}.


\subsection{Lattice Technicalities}

The general method of determining matrix elements is by
forming ratios of three-point to two-point correlation functions:
\begin{equation}
   R_{\alpha\beta}(t,\tau; \vec{p})
     =   {\langle N_\alpha(t;\vec{p}) {\cal O}(\tau; \vec{0})
         \overline{N}_\beta(0; \vec{p}) \rangle \over
         \langle N(t; \vec{p}) \overline{N}(0; \vec{p}) \rangle } \,,
\label{ratio}
\end{equation}
where $N$ is the three quark baryon operator and ${\cal O}$
is a bilinear quark operator. Provided that $t \gg \tau \gg 0$,
then $R$ can be shown to be directly proportional
to the matrix element $\langle N| {\cal O} |N \rangle$.
($R$ may be easily generalised to allow for momentum transfer.)
If we re-write the numerator of eq.~(\ref{ratio})
using quark propagators, we see that we have two classes of diagrams:
one in which the inserted operator is attached to quark propagators
to the nucleon, and a second class where the quark propagator
from the operator is disjoint from the baryon. In this second class the
interactions between the inserted operator and the baryon
take place only via gluon exchange. Numerically,
due to large ultra-violet fluctuations, no useful signal is seen.
So either we must presently assume that their effects 
are small, or calculate `non-singlet' (i.e. $u-d$) matrix elements
where these second class terms cancel.

The matrix elements must be renormalised. There are several possibilities.
We may simply consider a ratio of matrix elements between
different hadronic states, so that the renormalisation constant cancels.
However, this is not very useful. The axial and vector
renormalisation constants (the currents being partially conserved)
may be determined by demanding that their (continuum)
Ward Identities are obeyed, which produces non-perturbative
renormalisation constants. (Incidently they are also renormalisation
scheme independent.) Perturbation theory can be applied,
but due to technical problems only one loop results are known.
Even with this restriction, perturbation theory can be
improved leading to `tadpole improved' or {\it TI} perturbation theory.
However it still suffers from unknown systematic errors.
The ALPHA Collaboration has developed techniques\cite{sommer97a}
allowing some renormalisation constants to be determined
by stepping up from a low energy scale to a high energy scale
where the $Z$ can be matched to (known) perturbation theory.
Finally one can attempt to `mimic' the perturbation theory
method by calculating the operator numerically between quark states
in the Landau gauge. Also note that it is desirable
to make the errors in the determination to be of $O(a^2)$
rather than $O(a)$, because then the approach to the continuum
is faster\cite{luscher98a}. This introduces further irrelevant operators,
with `improvement' coefficients which also must be determined.

At present, for the axial and vector currents most
renormalisation constants and improvement coefficients
are known non-perturbatively. For $v_n$ we rely on 
{\it TI} perturbation theory.


\section{Results}


\subsection{The Axial Charge}

\begin{figure}[htb]
   \hspace*{0.75in}
   \epsfxsize=7.00cm \epsfbox{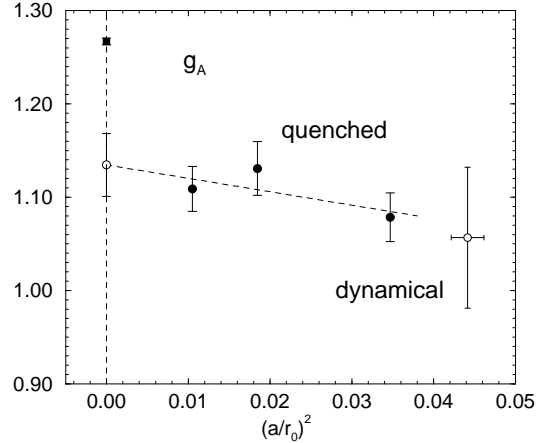}
   \caption{The continuum extrapolation of $g_A$.}
   \label{fig_ga_aor02_p0+dyn_ade01}
\end{figure}

Using the quenched approximation we have made simulations
at three lattice spacings: $a=0.093, 0.068$ and $0.051$ $\mbox{fm}$%
\footnote{Corresponding to $a^{-1} =2.12, 2.90$ and $3.85\,\mbox{GeV}$, 
with scale\cite{guagnelli98a} $r_0 = 0.5\,\mbox{fm}$.}.
For each lattice size simulations are made at four or more quark masses,
which are then linearly extrapolated to the chiral limit.
(A picture for $v_n$ will be shown later in section~\ref{section_moments}.)
We also have one unquenched result at $a \approx 0.11\,\mbox{fm}$.
In Fig.~\ref{fig_ga_aor02_p0+dyn_ade01} we show all these numbers%
\cite{horsley00a}.
The results all lie somewhat lower than the experimental value,
and at present there is no clear difference between quenched and
unquenched fermions. (Note that for the quenched results,
this is a fully non-perturbative calculation; for the unquenched
result, there is still a little uncertainty.)
While it is possible to extrapolate
the quenched data to the continuum limit (although it would be very
desirable to have more points because the noise is uncomfortably large
at present), we see that we are only at the beginning of obtaining
results with dynamical fermions.


\subsection{Form Factors}

\begin{figure}[htb]
   \hspace*{0.75in}
   \epsfxsize=7.00cm \epsfbox{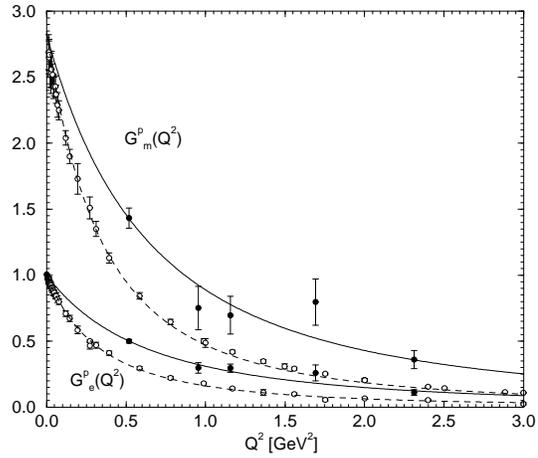}
   \caption{$G^p_e$ and $G^p_m$ compared to experimental data
            for quenched fermions at $a = 0.068\mbox{fm}$.}
   \label{fig_Ge+Gm_Vff_Qu-Qd_GeV2_b6p20_ade01}
\end{figure}

\begin{figure}[htb]
   \hspace*{0.75in}
   \epsfxsize=8.00cm \epsfbox{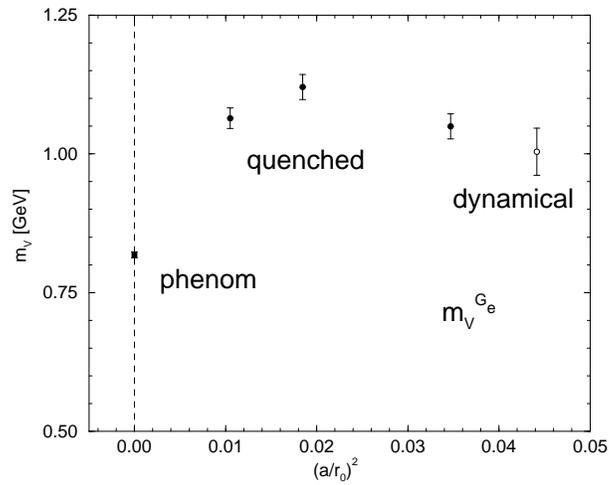} 
   \caption{$m_V^{G_e}$ compared to its phenomenological value.}
   \label{fig_Ge_p.fitparms_001018}
\end{figure}

In Fig.~\ref{fig_Ge+Gm_Vff_Qu-Qd_GeV2_b6p20_ade01} we show proton
electric and magnetic form factors\cite{capitani98a},
for our quenched lattice $a = 0.068\mbox{fm}$. Other lattice values
and the unquenched results are very similar.
Due to the momentum transfer the results are rather noisy. However
the present trend is clear: the dipole fit gives values of $m_V$
too large in comparison with the phenomenological value.
This is confirmed in Fig.~\ref{fig_Ge_p.fitparms_001018}.
For the nucleon we find $r_{\rm rms} \equiv \sqrt{12} / m_V \approx
0.60 \, \mbox{fm}$ in the quenched approximation and $r_{\rm rms} \approx 
0.70 \, \mbox{fm}$ for the dynamical case. This is to be compared with the 
phenomenological value $r_{\rm rms} = 0.83 \, \mbox{fm}$. Perhaps the `pion
cloud' has not developed at our dynamical quark masses, which makes the 
nucleon appear to be smaller than it really is. $\mu^p$ is roughly consistent
with the phenomenological value (but with large errors),
while $m_V^{G_m}$ is again too large.

\begin{figure}[htb]
   \hspace*{0.75in}
   \epsfxsize=7.00cm \epsfbox{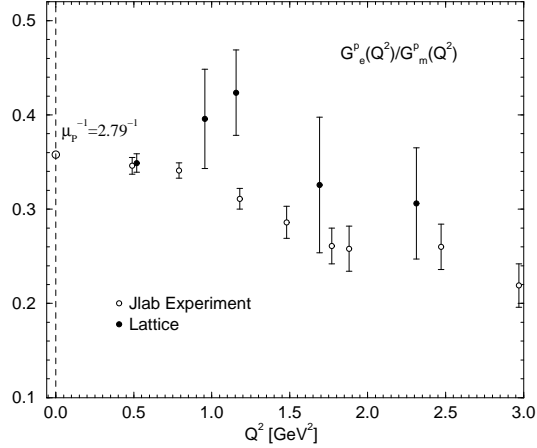}
   \caption{$G^p_e/G^p_m$ compared to experimental data
            for quenched fermions at $a = 0.068\mbox{fm}$.}
   \label{fig_Ge_o_Gm_b6p20_Vff_Qu-Qd_GeV2_ade01}
\end{figure}

The recent Jlab results\cite{jones99a} for $G^p_e/G^p_m$
indicate a functional difference between the electric
and magnetic form factors. This is shown in
Fig.~\ref{fig_Ge_o_Gm_b6p20_Vff_Qu-Qd_GeV2_ade01}.
If eq.~(\ref{Gem_phen_fits}) holds then the ratio of form factors
should be constant. Unfortunately at present the lattice data is too noisy
to draw any conclusion.

\begin{figure}[htb]
   \hspace*{0.75in}
   \epsfxsize=7.00cm \epsfbox{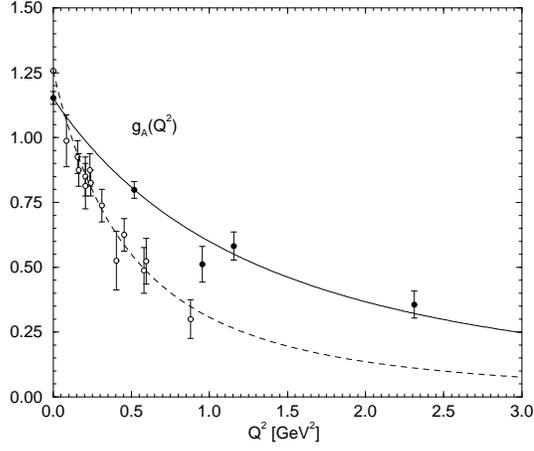}
   \caption{$g^p_A$ compared to experimental data
            for quenched fermions at $a = 0.068\mbox{fm}$.}
   \label{fig_GA_b6p20_Adr_1u-1d_GeV2_ade01}
\end{figure}

Finally in Fig.~\ref{fig_GA_b6p20_Adr_1u-1d_GeV2_ade01} we show
the axial form factor.
Again the previous comments as for $G^p_e$, $G^p_m$ hold verbatim.


\subsection{Moments of the nucleon structure function}
\label{section_moments}

\begin{figure}[htb]
   \hspace*{0.75in}
   \epsfxsize=7.00cm \epsfbox{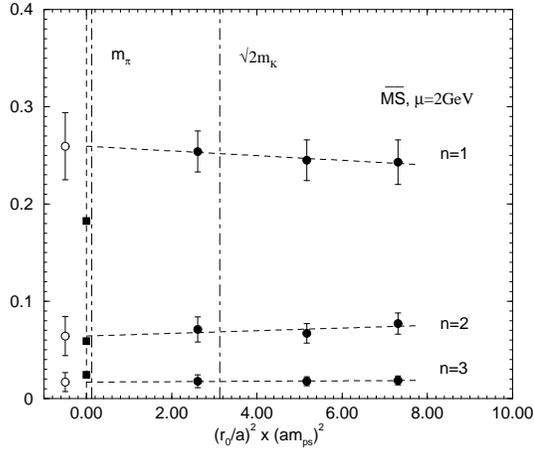}
   \caption{The chiral extrapolation for $v_n^{(u)\msbar}-v_n^{(d)\msbar}$ 
            for quenched fermions at $a = 0.093\mbox{fm}$ (circles),
            compared to the {\it MRS} fit function\protect\cite{martin95a}
            (filled squares).
            The dot-dashed lines show the positions of $m_\pi$
            and a (hypothetical) strange quark pseudoscalar particle.}
   \label{fig_xnu-d_wilson_ade01}
\end{figure}

Some years ago, we published our first results\cite{gockeler95a}
for $v_n$, $n = 1$, $2$, $3$. The re-plotted data for
$v_n^{({u})\msbar}-v_n^{(d)\msbar}$ is shown again in
Fig.~\ref{fig_xnu-d_wilson_ade01}.
for unimproved Wilson fermions. A comparison is made with
{\it MRS} phenomenological results\cite{martin95a}. While higher moments
tend to agree (although large error bars might hide
any discrepancy), it seems that there is a 
discrepancy for the lowest moment. This may be due to
renormalisation and discretisation effects, 
higher-twist contributions, the chiral extrapolation\cite{detmold01a},
or perhaps the phenomenological distribution functions, being {\it global} 
fits, do not fit this moment very well.
Naively this result is also perhaps expected as $v_2$
is part of the energy-momentum sum rule, $\sum v_2^{\rm valence} +
v_2^{\rm sea} =1$, and due to quenching the sea contribution is reduced.
For higher moments this is less of an effect, as the gluon
contribution is more important for smaller $x$.

\begin{figure}[t]
   \hspace*{0.75in}
   \epsfxsize=7.00cm \epsfbox{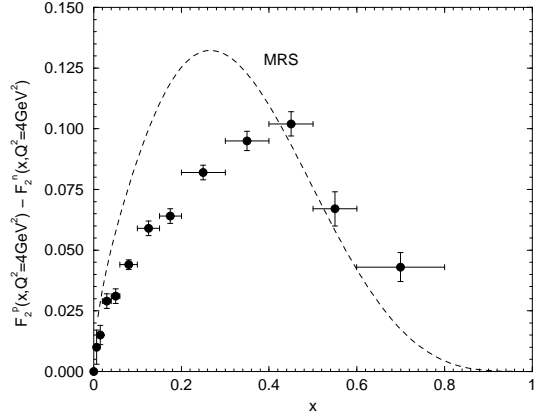}
   \caption{The experimental binned data for $F_2^p - F_2^n$
            at $Q^2 = 4\mbox{GeV}^2$ from the {\it NMC}
            Collaboration\protect\cite{arneodo94a} compared
            with the {\it MRS} parton fit function
            \protect\cite{martin95a} .}
   \label{fig_f2p-f2n}
\end{figure}

First let us look at the experimental data.
The cleanest {\it direct} determination of
$v_n^{(u)\msbar} - v_n^{(d)\msbar}$
is given from the $F_2^p - F_2^n$ {\it NMC} data\cite{arneodo94a},
as shown in Fig.~\ref{fig_f2p-f2n},
because the one experiment uses both $p$ {\it and} $n$ targets. Note
that there is a paucity of data at large $x \sim 1$, which is the
region where the moments are most sensitive to. (We make a simple
linear extrapolation from the largest data point to $x = 1$.)
While the {\it MRS} fit does not quite follow the data, the area
under the curve $\propto v_2^{(u)\msbar} - v_2^{(d)\msbar}$
is about the same, so there is indeed no problem.

\begin{figure}[tbh]
   \hspace*{0.75in}
   \epsfxsize=7.00cm \epsfbox{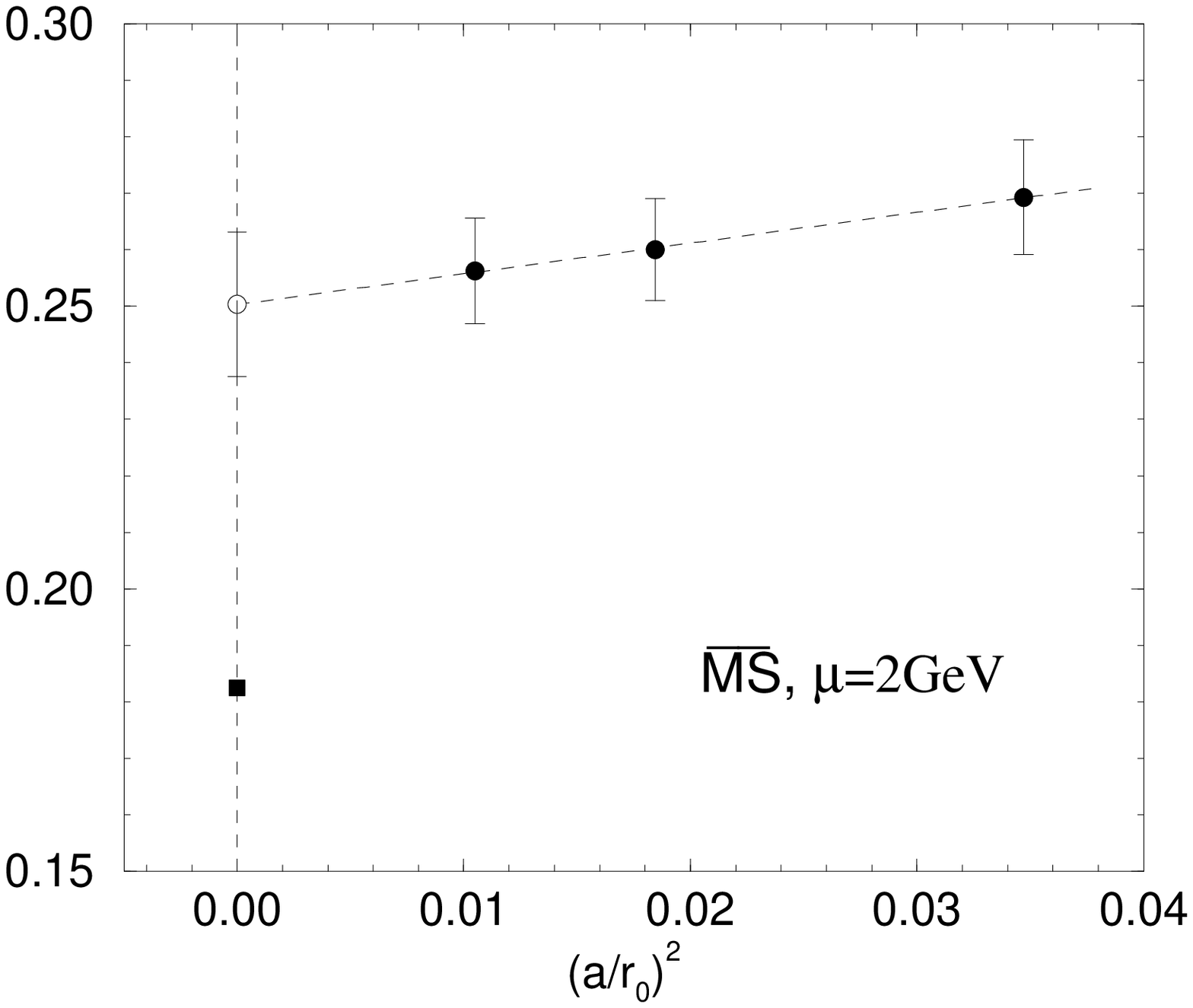}
   \caption{The continuum extrapolation for
            $v_2^{(u)\msbar}-v_2^{(d)\msbar}$.}
   \label{fig_x1b_aor02_p0_ade01}
\end{figure}

Secondly we have investigated the continuum extrapolation
for the lowest moment, in Fig.~\ref{fig_x1b_aor02_p0_ade01}
using improved fermions. Surprisingly, perhaps, there seems to be
little $a^2$ trend in the data (c.f. this with $g_A$,
Fig.~\ref{fig_ga_aor02_p0+dyn_ade01}). 

Thus the previous 
discrepancy still holds.


\section*{Acknowledgements}

The unquenched configurations were generated in collaboration
with {\it UKQCD} Collaboration. The propagators were computed 
using the {\it T3E} at ZIB and the {\it APE100} at NIC (Zeuthen).




\begin{thebibliography}{99}

\bibitem{bangalore00a} A good source of information is:
   Proceedings of the XVIIIth International Symposium on Lattice Field
   Theory, Bangalore, India, editors T. Bhattacharya, R. Gupta and A. Patel,
   {\it Nucl. Phys. Proc. Suppl.} {\underline{94}} (2001).

\bibitem{groom00a}
   Review of Particle Physics, D.~E. Groom et al, 
   {\it Eur. Phys. Jour.} {\underline{C15}} (2000) 1.

\bibitem{sommer97a}
   R. Sommer, Schladming Lectures (1997), hep-ph/9711243.

\bibitem{luscher98a}
   M. L\"uscher, Les Houches Lectures (1997), hep-ph/9711205.

\bibitem{guagnelli98a}
   M. Guagnelli, R. Sommer and H. Wittig,
   {\it Nucl. Phys.} {\underline{B535}} (1998) 389, hep-lat/9806005.

\bibitem{horsley00a}
   R. Horsley, Lattice 2000,
   {\it Nucl. Phys. Proc. Suppl.} {\underline{94}} (2001) 307,
   hep-lat/0010059.

\bibitem{capitani98a}
   S. Capitani, M. G\"ockeler, R. Horsley, B. Klaus, H. Oelrich,
   H. Perlt, D. Petters, D. Pleiter, P.~E.~L. Rakow, G. Schierholz,
   A. Schiller and P. Stephenson,
   {\it Nucl. Phys. Proc. Suppl.} {\underline{73}} (1999) 294,
   hep-lat/9809172.

\bibitem{jones99a}
   M.~K. Jones et al., 
   {\it Phys. Rev. Lett.} {\underline{84}} (2000) 1398, nucl-ex/9910005.

\bibitem{gockeler95a}
   M. G\"ockeler, R. Horsley, E.-M. Ilgenfritz, H. Perlt, P. Rakow,
   G. Schierholz and A. Schiller,
   {\it Phys. Rev.} {\underline{D53}} (1996) 2317, hep-lat/9508004.

\bibitem{martin95a}
   A. Martin, R.~G. Roberts and W.~J. Stirling,
   {\it Phys. Lett.} {\underline{B354}} (1995) 155, hep-ph/9502336.

\bibitem{detmold01a}
   W. Detmold, W. Melnitchouk, J.~W. Negele, D.~B. Renner
   and A.~W. Thomas, hep-lat/0103006.

\bibitem{arneodo94a}
   A. Arneodo et al., 
   {\it Phys. Rev.} {\underline{D50}} (1994) R1.

\end{thebibliography}
\end{document}